\documentclass[runningheads]{llncs}
\def\full{1}

\newcommand*{\cF}[1]{f^{-1}(#1)} 
\newcommand*{\cM}{\mathcal{M}}

\usepackage{amssymb}
\usepackage{amsmath}
\usepackage{amstext}
\usepackage{times}

\title{On the Impossibility of Extracting Classical Randomness Using a
Quantum Computer}

\author{Yevgeniy Dodis\inst{1} \and Renato Renner\inst{2}}

\institute{New York University, USA. Email: {\tt dodis@cs.nyu.edu}
 \and University of Cambridge, UK. Email: {\tt
 r.renner@damtp.cam.ac.uk}}

\date{}

\newcommand{\mypar}[1]{\vspace{2pt} \noindent {\sc #1}\ }

\newcommand*{\cE}{\mathcal{E}}
\newcommand*{\cH}{\mathcal{H}}
\newcommand*{\cP}{\mathcal{P}}
\newcommand*{\cU}{\mathcal{U}}
\newcommand*{\cX}{\mathcal{X}}
\newcommand*{\cY}{\mathcal{Y}}
\newcommand*{\cZ}{\mathcal{Z}}
\newcommand*{\Xp}{\tilde{X}}
\newcommand*{\Yp}{\tilde{Y}}

\newcommand*{\bcX}{\bar{\cX}}
\newcommand*{\bP}{\bar{P}}

\newcommand*{\ket}[1]{| #1 \rangle}
\newcommand*{\bra}[1]{\langle #1 |}

\DeclareMathOperator*{\tr}{tr}
\newcommand*{\id}{\mathrm{id}}

\newcommand*{\eps}{\varepsilon}

\newcommand{\ignore}[1]{}
\newcommand{\infull}[1]{\ifnum\full=1  {#1} \fi}

\begin{document}


\maketitle

\vspace{-2ex}
\begin{abstract}  
In this work we initiate the question of whether quantum computers can
provide us with an almost perfect source of classical randomness, and
more generally, suffice for classical cryptographic tasks, such as
encryption. Indeed, it was observed \cite{sv,mp90,DOPS04} that 
classical computers are insufficient for either one of these tasks
when all they have access to is a realistic {\em imperfect} source of
randomness, such as the Santha-Vazirani source.

\ignore{
On the other hand, quantum computation is inherently probabilistic
which suggests that perhaps quantum computers can provide a reasonable
way to overcome the above mentioned impossibility results.  In this
work
}
We answer this question in the {\em negative}, even in the following
very restrictive model.  We generously assume that quantum computation
is error-free, and {\em all the errors come in the measurements}.  We
further assume that all the measurement errors are not only small but
also {\em detectable}: namely, all that can happen is that with a
small probability $p_\perp\le \delta$ the (perfectly performed)
measurement will result in some distinguished symbol $\perp$
(indicating an ``erasure'').  Specifically, we assume that if an
element $x$ was supposed to be observed with probability $p_x$, in
reality it might be observed with probability $p_x'\in
[(1-\delta)p_x,p_x]$, for some small $\delta>0$ (so that $p_\perp = 1
- \sum_x p_x' \le \delta$).

\infull{
Our negative ``quantum'' result also implies a new ``classical''
result of independent interest: namely, even a much more restrictive
form of (classical) Santha-Vazirani sources is not sufficient for
randomness extraction and cryptography.
}

\ignore{
Quantum Computation is inherently noisy. The remarkable discovery of
quantum error-correcting codes has suggested that it is possible to
overcome this problem and design {\em fault-tolerant} quantum
computation. However, all the results in this area critically assumed
the noise is not only small, but essentially {\em independent}. This
situation can be parallelled with a similar study of classical
computation, where the initial enthusiasm of tolerating independent
errors --- going back to the pioneering work of von
Neumann~\cite{vN51} --- was somewhat lessened after several negative
realisations~\cite{SV85,MP91,DOPS04} came about pointing some of the
difficulties in dealing with general imperfect sources of randomness
}
\end{abstract}


\section{Introduction}

Randomness is important in many areas of computer science, such as
algorithms, cryptography and distributed computing. A common
abstraction typically used in these applications is that there exists
some source of unbiased and independent random bits.  However, in
practise this assumption seems to be problematic: although there seem
to be many ways to obtain somewhat random data, this data is almost
never uniformly random, its exact distribution is unknown, and,
correspondingly, various algorithms and protocols have to be based on
{\em imperfect sources of randomness}.

Not surprisingly, a large body of work (see below) has attempted to
bridge the gap between this convenient theoretical abstraction and the
actual reality. So far, however, most of this work concentrated on
studying if {\em classical} computers can effectively use classical
imperfect sources of randomness. In this work, we initiate the
corresponding study regarding {\em quantum computation}. To motivate
our question, we start by surveying the state of the art in using
classical computers, which will demonstrate that such computers are
provably incapable of tolerating even ``mildly'' imperfect random
sources.

\newcommand{\ab}{\allowbreak}

\mypar{Classical Approach to Imperfect Randomness.}  
The most straightforward approach to dealing with an imperfect random
source is to \emph{deterministically} (and efficiently) extract
nearly-perfect randomness from it. Indeed, many such results were
obtained for several classes of imperfect random sources. They include
various 
``streaming'' sources 
\ifnum\full=1
\cite{vN,elias,blum,lls},
\else
\cite{elias,blum,lls},
\fi
%
``bit-fixing'' sources
\ifnum\full=1
\cite{cgh,amplif,al,resilient,fourier,KZ03}, 
\else
\cite{cgh,al,fourier,KZ03}, 
\fi
multiple independent imperfect sources
\ifnum\full=1
\cite{sv,vazir,VazEff,cg,DO03,DEOR04,BIW04} 
\else
\cite{VazEff,cg,DO03,DEOR04,BIW04} 
\fi
and efficiently samplable sources \cite{tv}. While these results are
interesting and non-trivial, the above ``deterministically
extractable'' sources assume a lot of structure or independence in the
way they generate randomness.
A less restrictive, and arguably more realistic, assumption on the
random source would be to assume only that the source contains
\emph{some} entropy. We call such sources \emph{entropy sources}.
Entropy sources were first introduced by Santha and
Vazirani~\cite{sv}, and later generalised by Chor and
Goldreich~\cite{cg}, and Zuckerman~\cite{zuck}.

The entropy sources of Santha and Vazirani \cite{sv} are the least
imperfect (which means it is the hardest to show impossibility results
for such sources) among the entropy sources considered so far (e.g.,
as compared to \cite{cg,zuck}). SV sources, as they are called,
require {\em every bit output by the source} to have almost one bit of
entropy, even when conditioned on all the previous
bits. Unfortunately, already the original work of ~\cite{sv} (see also
a simpler proof in \cite{RVW:simplerproof}) showed that deterministic
randomness extraction of even a single bit is \emph{not} possible from
all SV sources. This can also be considered as impossibility of
pseudo-random generators with access to only an SV source. Moreover,
this result was later extended by McInnes and Pinkas~\cite{mp90}, who
showed that in the classical setting of computationally unlimited
adversaries, one cannot have secure symmetric encryption if the shared
key comes from an SV source.  Finally and most generally, Dodis et
al.~\cite{DOPS04} showed that SV sources in fact cannot be used
essentially for any interesting classical cryptographic task involving
privacy (such as encryption, commitment, zero-knowledge, multiparty
computation), even when restricting to computationally bounded
adversaries. Thus, even for the currently most restrictive entropy
sources, classical computation does not seem to suffice for
applications inherently requiring randomness (such as extraction and
cryptography).\footnote{In contrast, a series of celebrated positive
  results~\cite{vv85,sv,cg,zuck} show that even very weak entropy
  sources are enough for simulating probabilistic polynomial-time
  algorithms --- namely, the task which does not {\em inherently} need
  randomness. This result was extended to interactive protocols by
  \cite{DOPS04}. \cite{DOPS04} also show that under certain strong,
  but reasonable computational assumptions, secure signatures seem to
  be possible with entropy sources.}

We also mention that the impossibility results no longer hold when the
extracting party has a small amount of true randomness (this is the
study of so called probabilistic {\em randomness extractors}
\cite{nz}), or if several {\em independent} entropy sources are
available
\ifnum\full=1
\cite{sv,vazir,VazEff,cg,DO03,DEOR04,BIW04} 
\else
\cite{VazEff,cg,DO03,DEOR04,BIW04}.
\fi

\mypar{Quantum Computers?} Given the apparent inadequacy of classical
computers to deal with entropy sources --- at least for certain
important tasks such as cryptography ---, it is natural to ask if
quantum computers can be of help. 
More specifically, given that quantum computation is {\em inherently
probabilistic}, can we use quantum computers to generate nearly
perfect randomness? (Or maybe just ``good enough'' randomness for
cryptographic tasks like encryption, which, as we know \cite{DS02}, do
not require {\em perfect} randomness?)
For example, to generate a perfectly random bit from a fixed qubit
$\ket{0}$, one can simply apply the Hadamard transform, and then
measure the result in the standard basis. 
\ignore{
We notice that classical computation can be elegantly decoupled into a
way to generate a hopefully random input $r$, followed by a
deterministic procedure taking $r$ (and possibly the actual input
$x$). In contrast, quantum computation is {\em inherently
probabilistic}, without any need for some external ``randomness''
$r$. For example, to generate a perfectly random bit from a fixed
qubit $\ket{0}$, one can simply apply the Hadamard transform, and then
measure the result in the standard basis. This
suggests that perhaps the potential future existence of
quantum computers will easily resolve our need to generate nearly
uniform random bits. Indeed, we will not even have to rely on some
external physical source of randomness, since all such randomness can
be generated from ``the inside''.
}
Unfortunately, what prevents this simple solution from working in
practise is the fact that it is virtually impossible to perform the
above transformation (in particular, the measurement) precisely, so
the resulting bit is likely to be slightly biased. In other words, we
must deal with the noise.
\infull{
More generally, noise is a very serious issue in quantum computation,
which means that certain error-correction and fault-tolerance must be
applied in order to overcome such noise. Indeed, fault-tolerance is
one of the major problems in quantum computing (see \cite{NieChu00}),
so we will have to address it as well. Jumping ahead, however, what
will differentiate us from all the prior work in the area is the fact
that {\em we do not assume largely independent noise} (which can be
dealt with by quantum error-correction).
}

But, first, let us explain why there are good reasons to hope for
quantum computers to be useful despite the noise. When dealing with
classical imperfect sources, we usually assume that the source comes
from some family of distributions ``outside of our control'' (e.g.,
``nature''), so we would like to make as few assumptions about these
distributions as we can. For example, this is why the study of
imperfect randomness quickly converged to entropy sources as being the
most plausible sources one could get from nature. In contrast, by
using a quantum computer to generate our random source for us, we are
{\em proactively designing} a source of randomness which is convenient
for use, rather than {\em passively hoping} that nature will give us
such a source. Indeed, if not for the noise, it would be trivial to
generate ideal randomness in our setting. Moreover, even with noise we
have a lot of freedom in {\em adapting} our quantum computer to
generate and measure quantum states {\em of our choice}, depending on
the computation so far.

\mypar{Our Model.}  We first define a natural model for using a
(realistically noisy) quantum computer for the task of randomness
extraction (or, more generally, any probabilistic computation, such as
the one needed in classical cryptography). As we will see shortly, we
will prove a {\em negative} result in our model, despite the optimism
we expressed in the previous paragraph. Because of this, we will make
the noise as small and as restrictive as we can, even if these
restrictions are completely ``generous'' and unrealistic. Indeed, {\em
we will assume that the actual quantum computation is error-free, and
all the errors come in the measurements} (which are necessary to
extract some classical result out of the system). Of course, in
reality the quantum computation will also be quite noisy, but our
assumption will not only allow us to get a stronger result, but also
reduce our ``quantum'' question to a natural ``purely classical''
question of independent interest.

Moreover, we will further assume that all the measurement errors are
not only very small, but also {\em detectable}: namely, all that can
happen is that with a small probability $p_\perp\le \delta$ the
(perfectly performed) measurement will result in some distinguished
symbol $\perp$ (indicating an ``erasure'').  Specifically, we assume
that if an element $x$ was supposed to be observed with probability
$p_x$, in reality it might be observed with probability $p_x'\in
[(1-\delta)p_x,p_x]$, for some small $\delta>0$ (so that $p_\perp = 1
- \sum_x p_x' \le \delta$). Thus, it is guaranteed that no events of
small probability can be completely ``removed'', and the probability
of no event can be increased.  Moreover, as compared to the classical
SV model, in our model the state to be measured can be prepared
arbitrarily, irrespective of the computational complexity of preparing
this state. Further, such quantum states can even be generated
adaptively and based on the measurements so far. For comparison, in
the SV model the ``ideal'' measurement would always correspond to an
unbiased bit; additionally, the SV model allows for ``errors'' while
we only allow ``erasures''.

\mypar{Our Result.} Unfortunately, our main result will show that even
in this {\em extremely restrictive} noise model, one cannot extract
even a {\em single} nearly uniform bit. In other words, if the
measurement errors could be {\em correlated}, quantum computers do not
help to extract classical randomness. More generally, 
we extend the technique of \cite{DOPS04} to our model and show that
one cannot generate two (classical) computationally indistinguishable
distributions which are not nearly identical to begin with. This can
be used to show the impossibility of classical encryption, commitment,
zero-knowledge and other tasks exactly as in \cite{DOPS04}. We notice,
however, that our result does {\em not} exclude the possibility of
generating perfect entanglement, which might be used to encrypt a
message into a {\em quantum state}.  Nevertheless, our result implies
that, even with the help of such perfect entanglement, the user will
not be able to generate a (shared) classical key that can be used for
cryptographic tasks.  To summarise, we only rule out the possibility
of {\em classical} cryptography with {\em quantumly generated
randomness}, leaving open the question of (even modelling!) {\em
quantum} cryptography with noise.


Of independent interest, we reduce our ``quantum'' problem to the
study of a new classical source, which is considerably more
restrictive than the SV source (and this restriction can really be
enforced in our model). We then show a classical impossibility result
for our new source, which gives a non-trivial generalisation of the
corresponding impossibility result for the SV sources
\cite{DOPS04,sv}. From another angle, it also generalises the
impossibility of extraction from the so called ``bias-control
limited'' (BCL) sources of \cite{Dod01}.
As with our source, the most general BCL source considered in
\cite{Dod01} can adaptively generate samples from arbitrary
distributions (and not just random bits). However, the attacker is
given significantly more freedom in biasing the ``real''
distributions. First, all expected ``real'' distributions can be
changed to arbitrary statistically close ones (which gives more power
than performing ``detectable erasures''), and, second, a small number
of ``real'' distributions can be changed arbitrarily (which we do not
allow at all).

\ignore{

a BCL source can adaptively generate samples from
arbitrary distributions (and not just random bits). However, the
attacker is given more freedom in biasing the ``real'' distributions:
all distributions can be changed to arbitrary statistically close ones
(as opposed to ``small detectable erasures''), and a small number can
be changed arbitrarily (which we do not allow at all).
}

To summarise, our main results can be viewed in three areas:

\vspace{-1ex}
\begin{enumerate}
\item A model of using noisy quantum computers for classical
probabilistic computation.
\item A reduction from a ``quantum'' question to the classical
  question concerning a much more restrictive variant of the SV
  (or general BCL) source(s).
\item A non-trivial impossibility result for the classical source we
  define.
\end{enumerate}

\mypar{Relation to Quantum Error-Correction.} 
What differentiates us from the usual model of quantum computation
with noise is the fact that our errors are not assumed independent. In
particular, conventional results on fault-tolerant quantum computation
(such as the {\em threshold theorem}; see \cite{NieChu00} for more
details) do not apply in our model (as is apparent from our negative
results).
From another perspective, our impossibility result is not just a
trivial application of the principle that one can always and without
loss of generality postpone all the measurements until the end (a
useful observation true in the ``perfect measurement'' case). For
example, if all the measurements are postponed to the end, then we
might observe a single ``useless'' $\bot$ symbol with non-trivial
probability $\delta$, while with many measurements we are bound to
observe a lot of ``useful'' non-$\bot$ symbols with probability
exponentially close to one.

Nevertheless, in our
model one can trivially simulate probabilistic 
algorithms computing deterministic outputs, just as was the case for
the classical
computation.
For example, here we actually {\em can}
postpone all the measurements until the end, and then either obtain an
error (with probability at most $\delta$ in which case the computation
can be repeated), or the desired result (with probability arbitrarily
close to $1-\delta$).
Of course, this ``positive'' result only holds because our noise model
was made unrealistically restrictive (since we proved an {\em
impossibility result}). Thus, it would be interesting to define a less
restrictive (and more realistic!) error model --- for example where
the actual quantum computation is not error-free --- and see if this
feasibility result would still hold.

\infull{
Finally, the problem of detection errors has been studied in the
context of non-locality testing~\cite{ClaHor74,MaSaSe83,Massar02},
which tries to experimentally prove the intriguing phenomenon that the
behaviour of certain distant but entangled particles cannot be explained
by classical randomness. These results are of the same flavor as our
impossibility result. Indeed, they show that, if certain detection
probability is too low, then the outputs might be chosen in a
malicious way such that the resulting statistics does not imply
non-locality. To our knowledge, this is the only result where some
impossibility is proved, based on the assumption that certain errors
occur.
}

\ignore{
Finally, the problem of detection errors has been studied in the
context of non-locality testing. It is one of the most intriguing
predictions of quantum mechanics that the behaviour of two distant but
entangled particles is non-local, that is, it cannot be explained by
any theory based only on classical randomness, as long as no
superluminal communication is assumed. However, to exclude such
theories experimentally, detectors with high efficiency are needed.
Otherwise, it is possible that the detection probability of the
particle pairs emerging from the analysers is biased, such that the
statistics looks non-local, even if the behaviour of the particles was
governed by purely classical
randomness~\cite{ClaHor74,MaSaSe83,Massar02}. These results are of the
same flavor as our impossibility result. Indeed, they show that, if
the detection probability is too low, then the outputs might be chosen
in such a malicious way that the resulting statistics is not useful,
namely it does not imply non-locality. To our knowledge, this is the
only result where some impossibility is proved, based on the
assumption that certain errors occur.
}

\section{Definition of the source} \label{sec:def}

A source with $n$ outputs $X_1, X_2, \ldots, X_n$ is specified by a
joint probability distribution $P_{X_1 \cdots X_n}$. However, for most
realistic sources, the actual distribution $P_{X_1 \cdots X_n}$ can
usually not be fully determined. Instead, only a few characteristics
of the source are known, e.g., that the conditional probability
distributions\footnote{We write $X^{k}$ to denote the $k$-tuple $(X_1,
  \ldots, X_k)$.} $P_{X_i|X^{i-1}}$ have certain properties. A
well-known example for such a characterisation are the Santha-Vazirani
sources.

\begin{definition}[\cite{sv}]
  A probability distribution $P_{X_1 \cdots X_n}$ on $\{0,1\}^n$ is an
  \emph{$\alpha$-SV source} if\footnote{$P_{X_i|X^{i-1}=x^{i-1}}$
    denotes the probability distribution of $X_i$ conditioned on the
    event that the $(i-1)$-tuple $X^{i-1} = (X_1, \ldots, X_{i-1})$
    takes the value $x^{i-1} = (x_1, \ldots, x_{i-1})$.}
  for all $i \in \{1, \ldots, n\}$ and $x^{i-1} \in \{0,1\}^{i-1}$ we have
  \[
    P_{X_i|X^{i-1}=x^{i-1}}(0) \in [\alpha,1-\alpha]
  \]
\end{definition}

We will define a more general class of sources which, in some sense,
includes the SV sources
\infull{(cf.\ Appendix).} 
The main motivation for our definition is to capture any kind of
randomness that can be generated using imperfect (quantum) physical
devices.  Indeed, we will show in Section~\ref{sec:quantum} that the
randomness generated by any imperfect physical device cannot be more
useful than the randomness obtained from a source as defined below.

Intuitively, a source can be seen as a device which sequentially
outputs symbols $X_1, \ldots, X_n$ from some alphabet $\cX$. Each
output $X_i$ is chosen according to some fixed probability
distribution which might depend on all previous outputs $X_1, \ldots,
X_{i-1}$.  The ``imperfectness'' of the source is then modelled as
follows. Each output $X_i$ is ``erased'' with some probability
$p_\perp$, i.e., it is replaced by some distinguished symbol $\perp$.
This erasure probability might depend on the actual output $X_i$ as
well as on all previous outputs $X_1, \ldots, X_{i-1}$, but is upper
bounded by some fixed parameter $\delta$.

Before stating the formal definition, let us introduce some notation
to be used in the sequel. For any set $\cX$, we denote by $\bcX$ the
set $\bcX := \cX \cup \{\perp\}$ which contains an extra symbol
$\perp$. For a probability distribution $P_X$ on $\cX$ and $\delta
\geq 0$, let $\cP^\delta(P_X)$ be the set of probability distributions
$\bP_X$ on $\bcX$ such that
\[
  (1-\delta) P_X(x) \leq \bP_X(x) \leq P_X(x) \ , 
\]
for all $x \in \cX$. In particular, the probability of the symbol
$\perp$ is bounded by $\delta$, that is, $\bP_X(\perp) \leq \delta$.

\begin{definition}
  Let $\delta \geq 0$ and let, for any $i \in \{1, \ldots, n\}$,
  $Q_{X_i|X^{i-1}}$ be a channel\footnote{A \emph{channel} $Q_{Y|X}$
    from $\cX$ to $\cY$ is a function on $\cY \times \cX$ such that,
    for any $x \in \cX$, $Q_{Y|X=x} := Q_{Y|X}(\cdot ,x)$ is a
    probability distribution on $\cY$.} from $\bcX^{i-1}$ to $\cX$.  A
  probability distribution $P_{X_1 \cdots X_n}$ on $\bcX^n$ is a
  \emph{$(\delta, \{Q_{X_i|X^{i-1}}\})$-source} if
  for all $i \in \{1, \ldots, n\}$ and $x^{i-1} = (x_1, \ldots,
  x_{i-1}) \in \bcX^{i-1}$ we have
  \[
    P_{X_i|X^{i-1} = x^{i-1}} \in \cP^\delta(Q_{X_i|X^{i-1}=x^{i-1}}) 
  \]
\end{definition}

In \infull{the Appendix}, we show that $(\delta,
\{Q_{X_i|X^{i-1}}\})$-sources can be used to simulate $\alpha$-SV
sources, for some appropriately chosen $\alpha$.  This means that
$(\delta, \{Q_{X_i|X^{i-1}}\})$-sources are at least as useful as SV
sources. The other direction is, however, not true.  That is,
$(\delta, \{Q_{X_i|X^{i-1}}\})$-sources have a strictly
less``malicious'' behaviour than SV sources (which makes our
impossibility proofs stronger).

\section{The quantum model} \label{sec:quantum}

In this section, we propose a model that describes the extraction of
classical information from imperfect quantum physical devices.
Clearly, our considerations also include purely classical systems as a
special case.

First, in Section~\ref{sec:perf}, we review the situation where the
quantum device is perfect. In this case, the process of extracting
randomness can most generally be seen as a sequence of perfect quantum
operations and perfect measurements. Then, in
Section~\ref{sec:imperf}, we consider the imperfect case where the
quantum device is subject to (malicious) noise. As we shall see, in
order to get strong impossibility results, it is sufficient to extend
the standard notion of perfect measurements by the possibility of
detectable failures in the measurement process.

\subsection{The perfect case} \label{sec:perf}

Let us briefly review some basic facts about quantum mechanics. The
\emph{state} of a quantum system is specified by a projector
$P_{\ket{\psi}}$ onto a vector $\ket{\psi}$ in a Hilbert space $\cH$.
More generally, if a system is prepared by choosing a state from some
family $\{\ket{\psi_z}\}_{z \in \cZ}$ according to a probability
distribution $P_Z$ on $\cZ$, then the behaviour of the system is fully
described by the \emph{density operator} $\rho := \sum_{z \in \cZ}
P_Z(z) P_{\ket{\psi_z}}$.  The most general \emph{operation} that can
be applied on a quantum system is specified by a family $\cE =
\{E_x\}_{x \in \cX}$ of operators on $\cH$ such that $\sum_{x \in \cX}
E_x^\dagger E_x = \id_{\cH}$ (see, e.g.,~\cite{NieChu00}). When $\cE$
is applied to a system which is in state $\rho$, then, with
probability $P_X(x) := \tr(E_x \rho E_x^\dagger)$, the classical
output $x \in \cX$ is produced and the final state $\rho_x$ of the
system is $\rho_x := \frac{1}{P_X(x)} E_x \rho E_x^\dagger$.  Hence,
when ignoring the classical output $x$, the state $\cE(\rho)$ of the
system after applying the operation $\cE$ is the average of the states
$\rho_x$, that is, $\cE(\rho) := \sum_{x} P_X(x) \rho_x = \sum_{x} E_x
\rho E_x^\dagger$.

It is important to note that also the action of preparing a quantum
system to be in a certain state $\rho_0$ can be described by a quantum
operation $\cE$. To see this, let $\rho_0$ be given by $\rho_0 =
\sum_{z \in \cZ} P_Z(z) P_{\ket{\psi_z}}$, for some family of vectors
$\{\ket{\psi_z}\}_{z \in \cZ}$ and a probability distribution $P_Z$ on
$\cZ$.  Additionally, let $\{\ket{i}\}_{i \in \{1, \ldots, d\}}$ be an
orthonormal basis of $\cH$. It is easy to verify that the quantum
operation $\cE = \{E_{z,i}\}_{z \in \cZ, i \in \{1, \ldots, d\}}$
defined by the operators
\[
E_{z,i} := \sqrt{P_X(z)} \ket{\psi_z} \bra{i}
\]
maps any arbitrary state $\rho$ to $\rho_0$, that is, $\cE(\rho) =
\rho_0$. 

We are now ready to describe the process of randomness extraction from
a quantum system. Consider a classical user with access to a quantum
physical device.  The most general thing he can do is to subsequently
apply quantum operations, where each of these operations provides him
with classical information which he might use to select the next
operation.  To describe this on a formal level, let $\cH$ be a Hilbert
space and let $\cX$ be a set. The strategy of the user in each step
$i$ is then defined by the quantum operation $\cE^{x^{i-1}} =
\{E^{x^{i-1}}_{x}\}_{x \in \cX}$ he applies depending on the classical
outputs $x^{i-1} \in \cX^{i-1}$ obtained in the previous steps.  Note
that, according to the above discussion, this description also
includes the action of preparing (parts of) the quantum system in a
certain state. We can thus assume without loss of generality that the
initial state of the system is given by some fixed projector
$P_{\ket{\psi_0}}$.  The probability distribution
$P_{X_i|X^{i-1}=x^{i-1}}$ of the classical outcomes in the $i$th step
conditioned on the previous outputs $x^{i-1}$ as well as the quantum
state $\rho_{x^{i}}$ after the $i$th step given the outputs $x^i$ is
then recursively defined by $\rho_{x^0} := P_{\ket{\psi_0}}$ and
\begin{align}
  P_{X_i|X^{i-1}=x^{i-1}}(x) \label{eq:probtrans}
& := 
  \tr(E^{x^{i-1}}_x \rho_{x^{i-1}} E_x^{x^{i-1}\dagger}) 
\\
  \rho_{x^{i}} = \rho_{(x^{i-1}, x)} \label{eq:statetrans}
& := 
  \frac{1}{P_{X_i|X^{i-1}=x^{i-1}}(x)} E_x^{x^{i-1}} \rho_{x^{i-1}} E_x^{x^{i-1}\dagger} \ .
\end{align}



\subsection{Quantum measurements with malicious noise} \label{sec:imperf}

We will now extend the model of the previous section to include
situations where the quantum operations are subject to noise. As we
are interested in proving the \emph{impossibility} of certain tasks in
the presence of noise, our results are stronger if we assume that only
parts of the quantum operation are noisy. In particular, we will
restrict to systems where only the classical measurements are subject
to perturbations.%
\footnote{To see that our model leads to strong impossibility results,
  consider for example an adversary who is allowed to transform the
  quantum state $\rho$ of the device into a state $\rho'$ which has at
  most trace distance $\delta$ to the original state $\rho$. Let $\cM$
  be a fixed von Neumann measurement and let $P$ be the distribution
  resulting from applying $\cM$ to $\rho$. It is easy to see that, for
  any given probability distribution $P'$ which is $\delta$-close to
  $P$, the adversary can set the device into a state $\rho'$ such that
  a measurement $\cM$ of $\rho'$ gives raise to the distribution $P'$.
  Consequently, such an adversary is at least as powerful as an
  adversary who can only modify the distribution of the measurement
  outcomes, as proposed in our model.  In particular, our
  impossibility results also apply to this case.}

Formally, we define an imperfect quantum device by its behaviour when
applying any operation $\cE$. Let $\delta \geq 0$ and let $\cE =
\{E_{x,u}\}_{x \in \cX, u \in \cU}$ be a quantum operation which
produces two classical outcomes $x$ and $u$, where $x$ is the part of
the output that is observed by the user.  The operation $\cE$ acts on
the imperfect device as it would in the perfect case, except that each
output $x$ is, with some probability $\lambda_x \leq \delta$, replaced
by a symbol $\perp$, indicating that something went wrong.
Additionally, we assume that, whenever such an error occurs, the state
of the system remains unchanged.\footnote{This means that, even if a
  measurement error occurs, the state of the quantum system is not
  destroyed. (Recall that our impossibility results are stronger the
  closer our model is to a model describing perfect systems.)} The
resulting probability distribution $P_{X}$ of the outputs when
applying $\cE$ to an imperfect device in state $\rho$ is thus given by
\[
  P_{X}(x) 
:= 
  \sum_{u} (1-\lambda_{x}) \tr(E_{x,u} \rho E_{x,u}^\dagger) \ .
\]
In particular, the probability of the symbol $\perp$ is $P_X(\perp) =
1-\sum_{x \in \cX} P_X(x) \leq \delta$.
    
Let us now consider the interaction of a user with such an imperfect
quantum device. In each step $i$, he either observes the correct
outcome or he gets the output $\perp$, indicating that something went
wrong.  The user might want to use this information to choose the
subsequent operations.  His strategy is thus defined by a family
$\{\cE^{x^{i-1}}\}_{x^{i-1} \in \bcX^{i-1}}$ of quantum operations
$\cE^{x^{i-1}} = \{E^{x^{i-1}}_{x,u}\}_{x \in \cX, u \in
  \cU}$.\infull{\footnote{Note that, unlike in the perfect case, the
    measurements cannot be postponed to the end of the protocol. For
    example, if the user performs many measurements during the
    protocol, it is very unlikely that \emph{all} the outcomes are
    wrong, i.e., he still gets some useful information with
    probability almost one. On the other hand, if the user replaces
    all his measurements by \emph{one} single overall measurement (at
    the end of the protocol) it might fail with
    probability~$\delta$.}}  The conditional probability distributions
$P_{X_i|X^{i-1}=x^{i-1}}$ of the observed outputs in the $i$th step,
for $x^{i-1} \in \bcX^{i-1}$, and the states $\rho_{x^i}$ after the
$i$th step are recursively defined, analogously
to~\eqref{eq:probtrans} and~\eqref{eq:statetrans}, by
\begin{align*}
  P_{X_i|X^{i-1}=x^{i-1}}(x) 
& :=
  (1-\lambda_{x^{i-1},x}) Q_{X_i|X^{i-1}=x^{i-1}}(x) \qquad \text{for $x \in \cX$}
\\
  \rho_{x^{i}} = \rho_{(x^{i-1}, x)}
& := 
\begin{cases}
  \frac{1}{Q_{X_i|X^{i-1}=x^{i-1}}(x)}
  \sum_{u \in \cU} 
    E_{x,u}^{x^{i-1}} \rho_{x^{i-1}} E_{x,u}^{x^{i-1}\dagger} &
  \text{if $x \in \cX$} \\
  \rho_{x^{i-1}} & \text{if $x = \perp$} \ .
\end{cases}
\end{align*}
for some $\lambda_{x^{i-1},x} \in [0,\delta]$, where $Q_{X_i|X^{i-1}}$
is the channel from $\bcX^{i-1}$ to $\cX$ given by
$Q_{X_i|X^{i-1}=x^{i-1}}(x) := \sum_{u \in \cU} \tr(E_{x,u}^{x^{i-1}}
\rho_{x^{i-1}} E_{x,u}^{x^{i-1}\dagger})$.

Let $P_{X^n} = P_{X_1 \cdots X_n}$ be the probability distribution of
the observed outcomes after $n$ steps. It follows directly from the
above formulas that $P_{X^n}$ is a $(\delta,
\{Q_{X_i|X^{i-1}}\})$-source.  On the other hand, if $P_{X^n}$ is a
$(\delta, \{Q_{X_i|X^{i-1}})$-source, then there exist weights
$\lambda_{x^{i-1},x} \in [0,\delta]$ such that the conditional
probabilities are given by the above formulas. This reduces our
``quantum'' problem to a totally classical problem for an imperfect
source considerably more restrictive than an SV source
\infull{(see Appendix).}
The corresponding impossibility result is given in the next section.

\section{Main technical lemma}

Our main technical result can be seen as an extension of a result
proved for SV sources (cf.\ Lemma~3.5 of~\cite{DOPS04}). Roughly
speaking, Lemma~\ref{lem:main} below states that a task $g$ which
requires perfect random bits can generally not be replaced by another
task $f$ which only uses imperfect bits. Note that this impossibility
is particularly interesting for cryptography where many tasks do in
fact use randomness.

More precisely, let $g$ be an arbitrary strategy which uses imperfect
randomness $X^n$ and, in addition, some perfect randomness $Y$ (whose
probability distribution might even
depend on the values of $X^n$). Let $f$ be another strategy which only
uses imperfect randomness $X^n$.  Furthermore, assume that, for any
$(\delta, \{Q_{X_i|X^{i-1}}\})$-source $P_{X_1 \cdots X_n}$, the
output distributions of the strategies $g$ and $f$ are (almost)
identical.  Then the strategy $g$ is (roughly) the same as $f$, that
is, it (virtually) does not use the randomness~$Y$.

\begin{lemma} \label{lem:main}
  Let $f$ be a function from $\bcX^n$ to $\cZ$, $g$ be a function from
  $\bcX^n \times \cY$ to $\cZ$ and $m =
  \lceil\log_2(|\cZ|)\rceil$. For any $i \in \{1, \ldots, n\}$, let
  $Q_{X_i|X^{i-1}}$ be a channel from $\bcX^{i-1}$ to $\cX$, let
  $Q_{Y|X^n}$ be a channel from $\bcX^n$ to $\cY$, and let $\delta
  \geq 0$.  Let $\Gamma$ be the set of all probability distributions
  $P_{X^n Y}$ on $\bcX^n \times \cY$ such that $P_{X^n}$ is a
  $(\delta, \{Q_{X_i|X^{i-1}}\})$-source\footnote{Similarly to the
  argument in~\cite{DOPS04}, the proof can easily be extended to a
  statement which holds for an even stronger type of sources, where
  the conditional probability distributions of each $X_i$ given all
  other source outputs, and not only the previous ones $X^{i-1}$, is
  contained in a certain set $\cP^{\delta}$.}  and $P_{Y|X^n} =
  Q_{Y|X^n}$.  If, for all $P_{X^n Y} \in \Gamma$,
  \[
    \|P_{f(X^n)} - P_{g(X^n, Y)}\|_1 < \eps \ ,
  \]
  then there exists $P_{\Xp^n \Yp} \in \Gamma$ such that
  \[
    \Pr_{(x^n, y) \leftarrow P_{\Xp^n \Yp}}[f(x^n) \neq g(x^n, y)]
  <
    5 \eps m \delta^{-1} \ ,
  \]
\end{lemma}

\begin{proof}  
  Assume first that the functions $f$ and $g$ are binary, i.e., $\cZ =
  \{0,1\}$. The idea is to define two probability distributions
  $P_{V^n Y}, P_{W^n Y} \in \Gamma$ such that the output distributions
  of the function $f$, $f(V^n)$ and $f(W^n)$, are ``maximally
  different''. Then, by assumption, the output distributions of
  $g(V^n, Y)$ and $g(W^n, Y)$ must be different as well. This will
  then be used to conclude that the outputs of $f$ and $g$ are
  actually equal for most inputs.
  
  In order to define the distributions $P_{V^n Y}$ and $P_{W^n Y}$, we
  first consider some ``intermediate distribution'' $P_{\Xp^n \Yp}$.
  It is defined as the unique probability distribution on $\bcX^n
  \times \cY$ such that $P_{\Yp|\Xp^n} = Q_{Y|X^n}$ and, for any $i
  \in \{1, \ldots, n\}$ and $x^{i-1} \in \bcX^{i-1}$,
  \[
    P_{\Xp_i|\Xp^{i-1}=x^{i-1}}(x) := \begin{cases} 
      (1-\frac{\delta}{2}) Q_{X_i|X^{i-1}=x^{i-1}}(x) 
        & \text{if $x \in \cX$} \\
      \frac{\delta}{2} & \text{if $x = \perp$} \ .
    \end{cases}
  \]
  Note that $P_{\Xp_i|\Xp^{i-1}=x^{i-1}} \in
  \cP^\delta(Q_{X_i|X^{i-1}=x^{i-1}})$, i.e., $P_{\Xp^n}$ is a
  $(\delta, \{Q_{X_i|X^{i-1}}\})$-source, and thus $P_{\Xp^n \Yp} \in
  \Gamma$.
  
  The distribution $P_{V^n}$ is now defined from $P_{\Xp^n}$ by
  raising the probabilities of all values\footnote{For $z \in
    \{0,1\}$, $\cF{z} := \{x \in \bcX^n: f(x) = z\}$ denotes the
    preimage of $z$ under the mapping $f$.}  $x^n \in \cF{0}$ that $f$
  maps to $0$ and lowering the probabilities of all $x^n \in \cF{1}$.
  Similarly, $P_{W^n}$ is defined by changing the probabilities of
  $P_{\Xp^n}$ in the other direction. For the formal definition, we
  assume without loss of generality that $P_{f(\Xp^n)}(0) \leq
  \frac{1}{2}$ and set $\alpha := P_{f(\Xp^n)}(0) / P_{f(\Xp^n)}(1)$,
  i.e., $\alpha \leq 1$.  $P_{V^n}$ and $P_{W^n}$ are then given by
  \begin{align*}
    P_{V^n}(x^n) 
  & := \begin{cases} 
      P_{\Xp^n}(x^n) (1+\tau) & \text{if $x^n \in \cF{0}$} \\
      P_{\Xp^n}(x^n) (1-\alpha \tau) & \text{if $x^n \in \cF{1}$}
    \end{cases} \\
    P_{W^n}(x^n) 
  & := 
    \begin{cases} 
      P_{\Xp^n}(x^n) (1-\tau) & \text{if $x^n \in \cF{0}$} \\
      P_{\Xp^n}(x^n) (1+\alpha \tau) & \text{if $x^n \in \cF{1}$} \ ,
    \end{cases}
  \end{align*}
  where $\tau:=\frac{\delta}{4}$. Because
  \[
  \begin{split}
    \sum_{x^n \in \bcX^n} P_{V^n}(x^n)
  & =
    \sum_{x^n \in \cF{0}} P_{X^n}(x^n) (1+\tau)
    + \sum_{x^n \in \cF{1}} P_{X^n}(x^n) (1-\alpha \tau) \\
  & =
    P_{f(X^n)}(0) (1+\tau) + P_{f(X^n)}(1)(1-\alpha \tau)
  =
    1 \ ,
  \end{split}
  \]
  $P_{V^n}$ and, similarly, $P_{W^n}$, is indeed a probability
  distribution.
  
  We claim that $P_{V^n}$ and $P_{W^n}$ are $(\delta,
  \{Q_{X_i|X^{i-1}}\})$-sources. To see this, note first that, for any
  $i \in \{1, \ldots n\}$ and $x^i \in \bcX^i$, $(1-\alpha \tau)
  P_{\Xp^i}(x^i) \leq P_{V^i}(x^i)$ and $P_{V^i}(x^i)\leq (1+\tau)
  P_{\Xp^i}(x^i)$.  Hence, for any $x \in \cX$ and $x^{i-1} \in
  \bcX^{i-1}$,
  \begin{multline*}
    P_{V_i|V^{i-1}=x^{i-1}}(x)
  =
    \frac{P_{V_i V^{i-1}}(x, x^{i-1})}
      {P_{V^{i-1}}(x^{i-1})} 
  \geq
    \frac{(1-\alpha\tau) P_{\Xp_i \Xp^{i-1}}(x, x^{i-1})}
      {(1+\tau) P_{\Xp^{i-1}}(x^{i-1})} \\
  =
    \frac{1-\alpha\tau}{1+\tau} P_{\Xp_i|\Xp^{i-1}=x^{i-1}}(x) 
  =
    \frac{1-\alpha\tau}{1+\tau} 
      (1-\textstyle\frac{\delta}{2}) Q_{X_i|X^{i-1}=x^{i-1}}(x) \ .
  \end{multline*}
   Because $\alpha \leq 1$, we have $P_{V_i|V^{i-1}=x^{i-1}}(x) \geq
  (1-\delta) Q_{X_i|X^{i-1}=x^{i-1}}(x)$. Similarly,
  \[
    P_{V_i|V^{i-1}=x^{i-1}}(x)
   \leq
    \frac{1+\tau}{1-\alpha\tau} P_{\Xp_i|\Xp^{i-1}=x^{i-1}}(x)
  =
   \frac{1+\tau}{1-\alpha\tau} (1-\textstyle\frac{\delta}{2}) 
     Q_{X_i|X^{i-1}=x^{i-1}}(x)
  \]
  which implies $P_{V_i|V^{i-1}=x^{i-1}}(x) \leq
  Q_{X_i|X^{i-1}=x^{i-1}}(x)$.  Combining these inequalities, we
  conclude $P_{V_i|V^{i-1}=x^{i-1}} \in
  \cP^\delta(Q_{X_i|X^{i-1}=x^{i-1}})$, i.e., $P_{V^n}$ is a $(\delta,
  \{Q_{X_i|X^{i-1}}\})$-source. A similar computation shows that also
  the distribution $P_{W^n}$ is a $(\delta,
  \{Q_{X_i|X^{i-1}}\})$-source. Consequently, the distributions
  $P_{V^n Y}$ and $P_{W^n Y}$ defined by $P_{Y|V^n} = Q_{Y|X^n}$ and
  $P_{Y|W^n} = Q_{Y|X^n}$, respectively, are contained in the set
  $\Gamma$.
  
  
  Next, we will analyse the behaviour of the function $g$ for inputs
  chosen according to $P_{V^n Y}$ and $P_{W^n Y}$, respectively, and
  compare it to $f$.  For this, let $q_{x^n}$ be the probability that,
  given some fixed $x^n \in \bcX^n$, the output of $g$ is zero, i.e.,
  $q_{x^n} := \Pr_{y \leftarrow Q_{Y|X^n={x^n}}}[g(x^n, y) = 0]$.
  Because $P_{\Yp|\Xp^n} = P_{Y|V^n} = P_{Y|W^n} = Q_{Y|X^n}$, we get
  \vspace{-0.7ex}
  \[
    q_{x^n} = P_{g(\Xp^n,\Yp)|\Xp^n=x^n}(0) = P_{g(V^n, Y)|V^n=x^n}(0) =
  P_{g(W^n, Y)|W^n=x^n}(0) \ .
  \]
  The probability that the output of $f$ is zero for the distributions
  $P_{V^n}$ and $P_{W^n}$ can then, obviously, be written as
  \begin{align*}
    P_{f(V^n)}(0) & = \sum_{x^n \in \cF{0}} P_{\Xp^n}(x^n) (1+\tau)  \\
    P_{f(W^n)}(0) & = \sum_{x^n \in \cF{0}} P_{\Xp^n}(x^n) (1-\tau) \ .
  \end{align*}
  Similarly, for $g$, we have
  \begin{align*}
    P_{g(V^n,Y)}(0)
  & =
    \sum_{x^n \in \cF{0}} P_{\Xp^n}(x^n) (1+\tau) q_{x^n}
    + \sum_{x^n \in \cF{1}} P_{\Xp^n}(x^n) (1-\alpha\tau) q_{x^n} 
\\
    P_{g(W^n,Y)}(0)
  & =
    \sum_{x^n \in \cF{0}} P_{\Xp^n}(x^n) (1-\tau) q_{x^n}
    + \sum_{x^n \in \cF{1}} P_{\Xp^n}(x^n) (1+\alpha\tau) q_{x^n} \ .
  \end{align*}
  By assumption of the lemma, because, $P_{V^n Y}$ and $P_{W^n Y}$ are
  contained in the set $\Gamma$, the output distributions of $f$ and
  $g$ must be close, that is, $|P_{f(V^n)}(0) - P_{g(V^n,Y)}(0)| <
  \frac{\eps}{2}$ and $|P_{f(W^n)}(0) - P_{g(W^n,Y)}(0)| <
  \frac{\eps}{2}$, and hence $(P_{f(V^n)}(0) - P_{g(V^n,Y)}(0)) -
  (P_{f(W^n)}(0) - P_{g(W^n,Y)}(0)) < \eps$.  Replacing these
  probabilities by the above expressions leads to
  \begin{equation} \label{eq:sumbound}
    \sum_{x^n \in \cF{0}} P_{\Xp^n}(x^n) 2 \tau (1-q_{x^n}) 
  + \sum_{x^n \in \cF{1}} P_{\Xp^n}(x^n) 2 \alpha \tau q_{x^n}
  <
    \eps \ .
  \end{equation}
  Note that this imposes some restrictions on the possible values of
  $q_{x^n}$. Roughly speaking, if $f$ maps a certain input $x^n$ to
  $0$, then the probability $1-q_{x^n}$ that $g$ maps $x^n$ to $1$
  must be small.  In fact, as we shall see, \eqref{eq:sumbound}
  implies a bound on the probability that the outputs of $f$ and $g$
  are different.
  
  With the definition $p_{z, w} := P_{f(\Xp^n) g(\Xp^n, Y)}(z,w)$, for
  $(z, w) \in \{0,1\}^2$ and using again the assumption of the lemma,
  \[
    |p_{0,1} - p_{1,0}|
  =
    |(p_{0,0} + p_{0,1}) - (p_{0,0} +p_{1,0})| 
  =
    |P_{f(\Xp^n)}(0) - P_{g(\Xp^n, \Yp)}(0)| 
  <
    \frac{\eps}{2} \ ,
  \]
  hence, 
  \begin{equation} \label{eq:pdiff}
    \Pr_{(x^n, y) \leftarrow P_{\Xp^n Y}}  
      [f(x^n) \neq g(x^n, y)]
  \leq
    p_{0,1} + p_{0,1} + |p_{1,0} - p_{0,1}|   
  <
    2 p_{0,1} + \frac{\eps}{2} \ .
  \end{equation}  
  Using~\eqref{eq:sumbound} and the fact that the second sum is
  nonnegative, we get an upper bound for $p_{0,1}$, that is,
  \[
  \begin{split}
    p_{0,1} 
  & =
    \sum_{x^n \in \bcX^n} P_{\Xp^n}(x^n) 
      P_{f(\bar{X^n})|\Xp^n=x^n}(0) 
      P_{g(\Xp^n, \Yp)|\Xp^n =x^n}(1) \\
  & =
    \sum_{x^n \in \cF{0}} P_{\Xp^n}(x^n) (1-q_{x^n}) 
  <
    \frac{\eps}{2 \tau} = \frac{2\eps}{\delta} \ .
  \end{split}
  \]
  Combining this with~\eqref{eq:pdiff}, we conclude $\Pr_{(x^n, y)
    \leftarrow P_{\Xp^n Y}} [f(x^n) \neq g(x^n, y)] <
  \frac{4\eps}{\delta} + \frac{\eps}{2} \leq \frac{5\eps}{\delta}$,
  which proves the lemma for the binary case where $\cZ = \{0,1\}$.
  
  To deduce the statement for arbitrary sets $\cZ$, consider an
  (injective) encoding function $c$ which maps each element $z \in
  \cZ$ to an $m$-tuple $(c_1(z), \ldots, c_m(z))$. Since the
  $L_1$-norm $\|\cdot\|_1$ can only decrease when applying a function,
  the assumption of the lemma implies that, for all probability
  distributions $P_{X^n Y} \in \Gamma$, $\|P_{f_k(X^n)} - P_{g_k(X^n,
    Y)}\|_1 < \eps$, where $f_k := c_k \circ f$ and $g_k := c_k \circ
  g$, for any $k \in \{1, \ldots, m\}$. The assertion then follows
  from the binary version of the lemma and the union bound. \qed
\end{proof}

As was shown in \cite{DOPS04}, Lemma \ref{lem:main}
implies not only impossibility of extracting nearly perfect
randomness, but also impossibility of doing almost any classical task
involving privacy (such as encryption, commitment, etc.). For
illustrative purposes, we give such an argument for extraction,
referring to \cite{DOPS04} regarding the other tasks.



\begin{corollary}
  Let $f$ be a function from $\bcX^n$ to $\{0,1\}$ and $P_U$ be the
  uniform distribution on $\{0,1\}$. For any $i \in \{1, \ldots,
  n\}$, let $Q_{X_i|X^{i-1}}$ be a channel from $\bcX^{i-1}$ to $\cX$,
  and let $\delta \geq 0$. Then there exists a $(\delta,
  \{Q_{X_i|X^{i-1}}\})$-source $P_{X^n}$ such that
  \[
    \|P_{f(X^n)} - P_U\|_1 \geq \frac{\delta}{10} \ ,
  \]
\end{corollary}

\begin{proof}
  Assume by contradiction that, for any $(\delta,
  \{Q_{X_i|X^{i-1}}\})$-source $P_{X^n}$, $\|P_{f(X^n)} - P_U\|_1 <
  \frac{\delta}{10}$.  Let $g$ be the function on $\cX^n \times
  \{0,1\}$ defined by $g(x^n, u):=u$. Then, for any probability
  distribution $P_{X^n U} = P_{X^n} \times P_U$, where $P_{X^n}$ is a
  $(\delta, \{Q_{X_i|X^{i-1}}\})$-source, we have $\|P_{f(X^n)} -
  P_{g(X^n, U)}\|_1 < \frac{\delta}{10}$.  Lemma~\ref{lem:main} thus
  implies that there exists a $(\delta, \{Q_{X_i|X^{i-1}}\})$-source
  $P_{\Xp^n}$ with $\Pr_{(x^n, u) \leftarrow P_{\Xp^n} \times
    P_U}[f(x^n) \neq g(x^n, u)] < \frac{1}{2}$, that is, $\Pr_{(x^n,
    u) \leftarrow P_{\Xp^n} \times P_U}[f(x^n) \neq u] < \frac{1}{2}$.
  This is a contradiction because $P_U$ is the uniform distribution on
  $\{0,1\}$. \qed
\end{proof}

\infull{
\appendix

\section*{Appendix: Relation to Santha-Vazirani sources} 

Let $P_{X^n}$ be a $(\delta, \{Q_{X_i|X^{i-1}}\})$-source, for some
$\delta \geq 0$ and channels $Q_{X_i|X^{i-1}}$. It is easy to verify
that, if $\delta \leq \frac{1}{|\cX|}$ then the entropy of the $i$th
output $X_i$ conditioned on any value of the previous outputs $X_1,
\ldots, X_{i-1}$ is lower bounded by the entropy of
$Q_{X_i|X^{i-1}=x^{i-1}}$, i.e.,
\begin{equation} \label{eq:entrinc}
  H(X_i|X^{i-1}=x^{i-1}) 
= 
  H(P_{X_i|X^{i-1} = x^{i-1}}) 
\geq 
  H(Q_{X_i|X^{i-1}=x^{i-1}}) \ ,
\end{equation}
for any $x^{i-1} \in \bcX^{i-1}$. This holds with respect to any
``reasonable'' entropy measure $H$, as, for instance, the Shannon
entropy, the min-entropy, or, more generally, the R\'enyi entropy of
order $\alpha$, for any $\alpha \in [0,\infty]$.

It is thus not surprising that $(\delta, \{Q_{X_i|X^{i-1}}\})$-sources
are at least as useful as Santha-Vazirani sources.  More precisely,
Lemma~\ref{lem:sv} below states that, for any $\alpha$, there exist
channels $Q_{X_i|X^{i-1}}$ and a deterministic\footnote{Note that any
  probabilistic strategy would require additional (perfect)
  randomness.} strategy $\gamma$ which allows to simulate an
$\alpha$-SV source from any $(\delta, \{Q_{X_i|X^{i-1}}\})$-source,
for $\delta = 1-2 \alpha$.  Hence, any impossibility result for
$(\delta, \{Q_{X_i|X^{i-1}}\})$-sources also holds for $\alpha$-SV
sources.

\begin{lemma}\label{lem:sv}
  For any $\delta \geq 0$, there exist channels $Q_{X_i|X^{i-1}}$, for
  $i \in \{1, \ldots, n\}$, and a function $\gamma$ such that the
  following holds: Let $P_{X^n}$ be an arbitrary $(\delta,
  \{Q_{X_i|X^{i-1}}\})$-source. Then the probability distribution
  $P_{Y^n}$ defined by $Y_i := \gamma(X_i)$, for $i \in \{1, \ldots
  n\}$, is an $\alpha$-SV source, for $\alpha = \frac{1-\delta}{2}$.
\end{lemma}

\begin{proof}
  Let $P_\delta$ be the binary probability distribution with
  $P_\delta(0) = \frac{1 + \delta}{2}$. For any $i \in \{1, \ldots,
  n\}$, let the channel $Q_{X_i|X^{i-1}}$ be defined by
  $Q_{X_i|X^{i-1}=x^{i-1}} := P_\delta$. Additionally, let $\gamma$ be
  the function on $\{0,1,\perp\}$ defined by
  \[
    \gamma(x) := \begin{cases} 
      x & \text{if $x \in \{0, 1\}$} \\
      1 & \text{if $x = \perp$} \ . 
    \end{cases}
  \]
  It is easy to verify that, for any $i \in \{1, \ldots, n\}$ and
  $x^{i-1} \in \{0,1,\perp\}^{i-1}$,
  \begin{align*}
    P_{\gamma(X_i)|X^{i-1} = x^{i-1}}(0)
  & \leq 
    P_\delta(0) = \frac{1+\delta}{2} = 1-\alpha 
\\
    P_{\gamma(X_i)|X^{i-1} = x^{i-1}}(0)
  & \geq
    P_\delta(0) (1-\delta) = \frac{1+\delta}{2} (1-\delta) \geq \alpha \ ,
  \end{align*}
  i.e., $P_{\gamma(X_i)|X^{i-1} = x^{i-1}}(0) \in [\alpha, 1-\alpha]$.
  By convexity, it follows that $P_{Y_i|Y^{i-1} = y^{i-1}}(0) \in
  [\alpha, 1-\alpha]$, for any $y^{i-1} \in \{0,1\}^{i-1}$, which
  concludes the proof. \qed
\end{proof}

Note that the converse of Lemma~\ref{lem:sv} is not true, i.e.,
Santha-Vazirani sources are generally weaker than $(\delta,
\{Q_{X_i|X^{i-1}}\})$-sources.  To see this, let, e.g., for any $i \in
\{1, \ldots, n\}$, $Q_{X_i|X^{i-1}}$ be the channel defined by the
uniform distribution over $\cX := \{0,1\}$, i.e.,
$Q_{X_i|X^{i-1}=x^{i-1}}(0) = \frac{1}{2}$, for all $x^{i-1} \in
\bcX^{i-1}$. It follows from~\eqref{eq:entrinc} that the entropy of
any $(\delta, \{Q_{X_i|X^{i-1}}\})$-source $P_{X_1 \cdots X_n}$ is at
least $n$, for any small enough $\delta \geq 0$.  On the other hand,
the entropy of an $\alpha$-SV source $P_{Y_1 \cdots Y_n}$, for any
$\alpha \neq \frac{1}{2}$, is generally smaller than $n$. As the
entropy of a random variable can only decrease when applying a
(deterministic) function, the values $(Y_1, \ldots, Y_n)$ cannot be
used to simulate $(X_1, \ldots, X_n)$.
}


\newcommand{\etalchar}[1]{$^{#1}$}


\end{document}